\documentclass[aps,prl,twocolumn,superscriptaddress,tightenlines]{revtex4-1}

\usepackage {graphicx}
\usepackage {amsmath}
\usepackage {amsfonts}
\usepackage {amssymb}
\usepackage {mathrsfs}
\usepackage {bm}

\usepackage {epstopdf}

\newcommand {\be}{\begin{eqnarray}}
\newcommand {\ee}{\end{eqnarray}}
  
\begin{document}

\title{Large $D-2$ Theory of Superconducting Fluctuations in a Magnetic Field
and Its Application to Iron-Pnictides}



\author{James M. Murray}
\affiliation{Department of Physics and Astronomy, Johns Hopkins University, Baltimore, Maryland 21218}
\author{Zlatko Te\v{s}anovi\'{c}}
\affiliation{Department of Physics and Astronomy, Johns Hopkins University, Baltimore, Maryland 21218}


\date{\today}

\begin{abstract}
A Ginzburg-Landau approach to fluctuations of a layered superconductor
in a magnetic field is used to show that the interlayer coupling 
can be incorporated within an
interacting {\em self-consistent} theory of a {\em single}
layer, in the limit of a large number of neighboring layers.  
The theory exhibits {\em two} phase transitions -- a vortex 
liquid-to-solid transition is
followed by a Bose-Einstein condensation into the Abrikosov lattice --
illustrating the essential role of interlayer coupling.
Using this theory, explicit expressions for magnetization,
specific heat, and fluctuation conductivity are derived.  
We compare our results with recent experimental data on the 
iron-pnictide superconductors.
\end{abstract}


\maketitle


The discovery of high-temperature superconductivity 
in iron-pnictides \cite{kamihara08,pnictidereview} 
has led to a renewed interest in the physics of layered 
compounds and the role of
superconducting fluctuations.  In older 
high-T$_c$ superconducting cuprates, due in large part
to their extreme anisotropy, the fluctuations have taken center stage, 
particularly in a magnetic field \cite{reviews}. At present,
a rather good understanding of such
fluctations is available in two-dimensional (2D) and three-dimensional (3D) systems.
However, the intermediate regime, where the interlayer coupling is too weak to
be ignored and yet not strong enough to render the system fully 3D, 
remains an important challenge.
Although most theoretical 
models of pnictides so far have focused on the 2D
nature of these materials \cite{raghu08, kuroki08, cvetkovic09, Chubukov, Bernevig}, 
experimental evidence frequently suggests a pronounced quasi 3D
behavior \cite{salem09, choi09}, especially within the 
so-called 122 family \cite{rotter08}. Thus, the iron-pnictides
apparently belong to this in-between regime.

In this Letter, we introduce a theoretical approach that allows for an explicit 
approximate solution to the problem of superconducting fluctuations in this
challenging intermediate situation. First, we show that the
Josephson coupling between superconducting layers in a magnetic field can be 
recast as a contribution to the effective ``on-site" Ginzburg-Landau (GL) free 
energy of a {\em single} layer, in the limit of a 
{\em large} number of neighboring layers.  The system 
is thus described by an {\em effective} 2D GL 
theory, which -- for practical purposes -- can be treated exactly,
by solving a set of non-linear, self-consistent equations, in combination with
a solution for the purely 2D case \cite{tesanovic92, tesanovic94, ullah91, ikeda}.
Second, we show that this theory -- unlike the 2D one --
possesses {\em two} phase transitions, reflecting the {\em crucial}
role of Josephson coupling.
Finally, we apply our theory to study fluctuation effects 
around the upper critical field $H_{c2}(T)$ and compare the results
to recent experimental data on the iron-pnictide superconductors.

We consider a general Josephson-coupled layered system, with an individual
layer described by the GL model.  The partition function is
\begin{eqnarray}
Z = \prod_i \int D(\bar{\psi}_i,\psi_i)e^{-S_0(i)- \sum_{j(i)} S_{\mathrm{int}}(i, j)},
\label{firsteq}
\end{eqnarray}
\noindent where $\psi_i\in {\rm LLL}$ is the fluctuating
GL order parameter in the $i$th layer;
LLL denotes the lowest Landau level for charge $2e$; 
and $j$ is summed over nearest neighbors of layer $i$.  
The corresponding action is 
\begin{eqnarray}
S_0 (i) = \frac{s}{T} \int d^2r \left( \alpha |\psi_i ({\bf r})|^2 + \frac{\beta}{2} |\psi_i ({\bf r})|^4 \right)~,
\end{eqnarray}
\noindent where $s$ is the distance between layers, and $\alpha = \alpha_0 (t-t_{c2}(h))$, where $t = T/T_c(0)$ and $h = H / H_{c2}(0)$ are the dimensionless temperature and magnetic field, respectively.  The interlayer 
portion of the GL-LLL action \eqref{firsteq} is \cite{footnoteLD}
\begin{eqnarray}
\label{s_int}
S_{\mathrm{int}}(i,j) =  - \frac{s}{T} \sum_{j=1}^d
\int d^2r \frac{\eta}{2} \left[ \bar{\psi}_i ({\bf r}) \psi_j ({\bf r}) + \bar{\psi}_j ({\bf r}) \psi_i ({\bf r}) \right] .
\end{eqnarray}

The goal now is to integrate out the Josephson-coupled portion
and obtain a partition function for the 0th layer that is entirely ``local,"
i.e. defined on a {\em single} layer.  As a first step, we assume 
that this can be done for the layers (denoted by $j$) that are adjacent 
to the 0th layer, i.e.\ that all couplings 
$S_{int}(j, j + \sigma)$, where $\sigma$ denotes 
all layers neighboring layer $j$ except for the 0th 
layer, can be integrated over, giving a 
correction to the ``on-site" action, so that $S_0(j) \rightarrow S_0'(j)$.  
(When the number of layers $j$ is very large
they decouple from each other, and we are left with a Bethe lattice, where 
each lattice ``site" is actually a 2D superconducting layer 
and the coordination number of the lattice is $d$. 
This is different from Ref.\ \cite{kollar05}, where each site is
a 0D quantum cluster.)  We obtain
\begin{eqnarray}
\begin{split}
\label{Z0}
Z(0) =  & \int D(\bar{\psi}_0,\psi_0)e^{-S_0 (0)} 
\\ & \times \prod_{j=1}^d \frac{1}{Z_0 (j)} \int  D(\bar{\psi}_j,\psi_j) e^{-S'_0(j) - S_{\mathrm{int}}(0,j)},
\end{split}
\end{eqnarray}
\noindent where $Z_0(j) = Z(j)|_{S_{int}=0}$. Expanding the interlayer term in \eqref{Z0}, and noting that only 
even terms in the expansion will survive the functional integration, yields
\begin{eqnarray}
\begin{split}
\label{zee}
\sum_n \frac{1}{(2n)!} \left( \frac{\eta s}{T} \sum^d_{j=1} 
\int d^2r \left[ \bar{\psi}_0 ({\bf r}) \psi_j({\bf r}) + \bar{\psi}_j ({\bf r}) \psi_0 ({\bf r}) \right]   \right) ^{2n}.
\end{split}
\end{eqnarray}
\noindent The terms 
that survive the functional integral are
of the form $(\bar{\psi}_0 \psi_0)^n \bar{\psi}_{j_1} \psi_{j_1} \dots \bar{\psi}_{j_n} \psi_{j_n}$.  In the $d\to\infty$ limit, the large majority of these terms 
has $j_1 \ne j_2 \ne \dots \ne j_n$.  There are $(2n)!$ of each terms of this type.  Since each involves $n$ pairs, and since there are $d$ possible pairs to choose from, the total number of all such terms (note that $j$'s are indistinguishable) 
is $(2n)! \binom{d}{n} \xrightarrow{d \rightarrow \infty} (2n)! d^n / n!$. 

Thus, in the large-$d$ limit \eqref{zee} turns into
\begin{eqnarray}
\begin{split}
\label{nsum}
& \sum_n  \frac{1}{(2n)!} \left( \frac{\eta s}{T} \right) ^{2n}  \frac{d^n (2n)!}{n!} 
\left( \int D(\bar{\psi}_j,\psi_j) e^{-S'_0(j)} \right)^{d-n}
\\ & \times \left( \int D(\bar{\psi}_j,\psi_j) e^{-S'_0(j)} \int d^2r \int d^2r' \bar{\psi}_0 \psi_0' \bar{\psi}_j \psi_j'  \right)^n,
\end{split}
\end{eqnarray}
\noindent where we have adopted the shorthand $\psi \equiv \psi({\bf r})$ and $\psi' \equiv \psi({\bf r}')$. This expression can now be inserted into Eq.\ \eqref{Z0}, where
the sum over $n$ can be re-exponentiated, giving
\begin{eqnarray}
\begin{split}
\label{Z1}
Z^{(1)}(0) = & \int D(\bar{\psi},\psi)e^{-S_0} 
\\ & \times \exp{\left[ \left( \frac{\tilde{\eta} s}{T} \right)^2  \int d^2r \int d^2 r' \bar{\psi} \psi'  \left< \bar{\psi}  \psi' \right> \right]}.
\end{split}
\end{eqnarray}
\noindent  The superscript in $Z^{(1)}(0)$ signifies that this is the leading term in a large-$d$ expansion.  Here we have defined $\tilde{\eta} \equiv \eta \sqrt{d}$ as the new interlayer coupling, which remains finite as $\eta \rightarrow 0$ and $d \rightarrow \infty$.  The $j$ index has been dropped, since all layers are equivalent and are no longer coupled.  The general correlation function is defined as
\begin{eqnarray}
\label{corr}
\left< \dots \right> \equiv \frac{ \int D(\bar{\psi},\psi) ( \dots ) e^{-S'_0} }
{ \int D(\bar{\psi},\psi) e^{-S'_0} }.
\end{eqnarray}
\noindent In the symmetric gauge, the correlation function in \eqref{Z1} is 
\begin{eqnarray}
\label{at}
\left< \bar{\psi}({\bf r}) \psi ({\bf r}') \right> = \frac{T}{2 \pi l^2 s \tilde \alpha} e^{-(|z|^2 + |z'|^2)/4 + \bar{z} z' / 2}.
\end{eqnarray}
\noindent where $z = (x + iy)/l$ is the complex coordinate within a single layer, $l = \sqrt{\phi_0 / 2 \pi H}$ is the magnetic length, and $\tilde \alpha$ is defined later.  The integral in Eq.\ \eqref{Z1} is thus
\begin{eqnarray}
\begin{split}
&\frac{1}{\tilde \alpha} \int d^2r \int d^2 r' \bar{\psi} ({\bf r}) \psi ({\bf r}')  e^{-(|z|^2 + |z'|^2)/4 + \bar{z} z' / 2}
\\ & = \frac{2 \pi l^2}{\tilde \alpha} \int d^2 r |\psi ({\bf r})|^2.
 \end{split}
\end{eqnarray}
\noindent 
The last equality follows from $\psi\in {\rm LLL}$. 

Following Ref.\ \cite{tesanovic94}, we make change of
variables ${ \psi ({\bf r}) = \Phi \prod_i (z-z_i) e^{-|z|^2 / 4} }$,
where $\{ z_i \}$ are the positions of
vortices.  The interaction of $\{ z_i \}$ is set by
$U^{-1} \equiv \sqrt{\left< \beta_A \right>}$, where
$\beta_A (\{ z_i \})\equiv \overline{|\psi |^4}/ \overline{|\psi |^2}^2 $
is the Abrikosov ratio for {\em arbitrary}
$\{ z_i \}$ ($\overline{\cdots}$ denotes a
spatial average). The partition function for the zeroth layer becomes
\begin{eqnarray}
\begin{split}
\label{z_phi}
& Z^{(1)}(0) =  \int d \Phi^* d \Phi \int dU  e^{N s(U)} e^{-S_{\mathrm{eff}}}
\\ & S_{\mathrm{eff}} = \frac{2 \pi l^2 s N}{T} \left( \alpha' |\Phi|^2 + \frac{\beta}{2 U^2} |\Phi|^4 \right) - N \mathrm{ln} ( 2 \pi l^2 s |\Phi|^2 )
\end{split}
\end{eqnarray}
\noindent
Here $N$ is the number of vortices $\{ z_i \}$
and $\alpha ' \equiv \alpha - \tilde{\eta}^2 / \tilde \alpha $. 
The entropy function $s(U)$ contains all the
effects of lateral correlations among vortices $\{ z_i \}$,
and knowledge of its exact form
is equivalent to the exact solution for the thermodynamics of a single layer
\cite{tesanovic94}.


In the thermodynamic limit $N\to \infty$, the saddle point method can be 
applied to integrals over $\Phi$ and $U$ in Eq.\ \eqref{z_phi}.  Minimizing
with respect to $|\Phi|^2$ gives
\begin{eqnarray}
\label{phi_min}
|\Phi_0|^2 = \frac{1}{2} \left[ -\frac{\alpha' U^2}{\beta} + \sqrt{\left( \frac{\alpha' U^2}{\beta} \right) ^2 + \frac{4 T U^2}{2 \pi l^2 s \beta}} \right].
\end{eqnarray} 
\noindent In order for this expression to be useful, 
we must determine the form of $\tilde \alpha$, as well as $U$.

From Eq.\ \eqref{at}, we have 
$\tilde \alpha ^{-1} = (2\pi l^2 s / T)\left< \bar{\psi} (0) \psi (0) \right>$.  
Using this along with Eqs.\ \eqref{corr} and \eqref{phi_min}, 
we obtain the following self-consistent expression for $\tilde \alpha$:
\begin{eqnarray}
\tilde \alpha = \alpha - \frac{\tilde{\eta}^2}{\tilde \alpha} + \frac{\beta T}{2 \pi l^2 s \tilde \alpha U^2} .
\end{eqnarray}
\noindent Solving this for $\tilde \alpha$, and substituting the result into our expression for $\alpha'$, we get
\begin{eqnarray}
\label{alpha_prime}
\alpha'  =  \alpha \left[ 1- \frac{2 \left( \frac{\tilde \eta}{\alpha} \right)^2 }{1 + \mathrm{sgn}(\alpha) \sqrt{1+\frac{2}{\alpha ^2} \left( \frac{\beta T}{\pi l^2 s U^2} - 2 \tilde{\eta} ^2 \right)} } \right].
\end{eqnarray}
\noindent  
In solving for this expression, we must assume
$\beta '\equiv\beta - 2 \tilde{\eta} ^2 ( \pi l^2 s U^2 / T)> 0$.  
$\beta ' < 0$ leads to
$\tilde \alpha < 0$, which is clearly unphysical.  
The implications of $\beta '\to 0^+$ at 
finite $T$ are important and are discussed shortly.
Eq.\ \eqref{alpha_prime} constitutes our main theoretical result, 
allowing us to describe the system of {\em coupled} layers 
with a 2D GL-LLL action, albeit with
$\alpha \rightarrow \alpha'$. Its innocent appearance notwithstanding,
the change $\alpha \rightarrow \alpha'$ actually entails an elaborate
self-consistent calculation to determine the ultimate dependence 
on $T$ and $H$.  
Note that the next term in the large-$d$ 
expansion -- arising from terms in \eqref{zee} 
with one index repeated four times -- modifies
the quartic term $\beta$ in the 2D GL action.
It is important to systematically incorporate such finite-$d$ corrections 
when addressing the details of interlayer
correlations in real materials.  

Evaluating Eq.\ \eqref{z_phi} at its
saddle point and using Eq.\ \eqref{phi_min}, we obtain for the free energy density
\begin{eqnarray}
\begin{split}
\label{eff}
\frac{F}{V} = & \frac{H T}{\phi_0 s} \Bigg[ \frac{1}{2} - \frac{1}{2} g^2 U^2 + \frac{1}{2} g U \sqrt{2 + g^2 U^2}
\\ & - \mathrm{ln} \left( -g U^2 + U \sqrt{2 + g^2 U^2}\right) - \frac{1}{2} \mathrm{ln} \frac{\pi l^2 s T}{\beta} \Bigg],
\end{split}
\end{eqnarray}
\noindent where $g \equiv \alpha ' \sqrt{2 \pi l^2 s / (2 \beta T)}$ can be expressed using Eq.\ \eqref{alpha_prime} as
\begin{eqnarray}
\label{g}
g = g_0 \frac{t -t_{c2}(h)}{\sqrt{h t}} \left[ 1- \frac{\frac{2}{ (t-t_{c2}(h))^2} \left( \frac{\tilde{\eta}}{\alpha_0} \right) ^2}{\mathrm{sgn}(\alpha) \Xi(h,t) +1} \right].
\end{eqnarray}
\noindent 
In the above equation 
$g_0 \equiv  \sqrt{s \phi_0 H_{c2} (0) / 16 \pi \kappa^2 T_c (0)}$,
\begin{eqnarray}
\Xi(h,t) \equiv \left[ 1 + \frac{4}{ (t-t_{c2}(h))^2} \left( \frac{h t}{2 g_0 ^2 U^2(g)} - \left( \frac{\tilde{\eta}}{\alpha_0} \right) ^2 \right) \right] ^{1/2},
\end{eqnarray}
\noindent 
and we used the GL result 
$\alpha_0 ^2 / \beta = H_{c2}^2 (0) / 8 \pi \kappa^2$ \cite{tinkham96}. 

$g(t,h)$ \eqref{g} is the scaling variable of our theory.
Since $\Xi(h,t)$ depends on $U (g)$, 
Eq.\ \eqref{g} has the form $g = g(U(g))$. $U(g)$ is the same as 
in a purely 2D problem, but there
$g(t,h) = g_0 (t-t_{c2}(h)) / \sqrt{h t}$, so the $t$ and $h$
dependencies in our case are very different. $U(g)$ follows from minimization
of \eqref{z_phi} and relies on knowledge of $s(U)$. Here we can turn the
problem around and exploit the fact that $\beta_A (g)$ 
interpolates between its high- and low-$T$ 
limits of 2 and $\beta_\Delta \equiv 1.159$, respectively.  In particular,
\begin{eqnarray}
\label{u_plain}
U(g) = 0.818 - 0.110\ \mathrm{tanh} \left( \frac{g + c_1}{c_2} \right),
\end{eqnarray}
\noindent 
suggested in Ref.\ \cite{tesanovic94}, where $c_1 = 1.60$ and $c_2 = 2.66$ 
from the fit to the Monte Carlo results of Ref. \cite{kato93}, 
yields a virtually exact solution for fluctuation 
thermodynamics \cite{footnote}.  
This expression for $U(g)$ can then be used to solve 
self-consistently for $g$ in Eq.\ \eqref{g}.

It is now clear that the divergence in Eq.\ \eqref{alpha_prime},
associated with $\beta '\to 0^+$ and 
$T \to T_{\Delta}=2 \pi l^2 \tilde{\eta}^2 s / (\beta \beta_\Delta)$, 
is endowed with special significance.  As $T$ is lowered
toward $T_{\Delta}$, $g \rightarrow - \infty$ (since $g \propto \alpha'$), and 
thus $U(g) \rightarrow 1/\sqrt{\beta_\Delta}$. Therefore, at {\em finite}
temperature $T_{\Delta}$ the system undergoes a Bose-Einstein condensation 
transition into the Abrikosov lattice state. In a purely 2D ($\tilde \eta = 0$) theory, such a transition could occur only at $T=0$.
Once $\tilde \eta \not= 0$, this transition moves to finite $T_{\Delta}$,
which, over a large portion of an $H-T$ phase diagram, is far below
the vortex liquid-solid transition taking 
place at $T_M$, defined by $g=g_M\sim -7$
\cite{footnote}. As $H\to 0$, both $T_{\Delta}$ and $T_M$ tend
into $T_{c2}(H)$. This echoes the phase diagram of layered
superconductors proposed in Ref. \cite{tesanovic94a}.

We now turn our attention to fluctuation 
thermodynamics \cite{tesanovic94,kogan,koshelev}.
The magnetization follows from
$4 \pi M = -(1 / V) \partial F / \partial H$, with $|\Phi_0|^2$ 
given in \eqref{phi_min}:
\begin{eqnarray}
\label{mag}
\frac{4 \pi \phi_0 s M}{T_c (0)} = g_0 \sqrt{h t} \left( g U^2 - U \sqrt{2 + g^2 U^2} \right).
\end{eqnarray}

\begin{figure}
\includegraphics[width=0.4\textwidth]{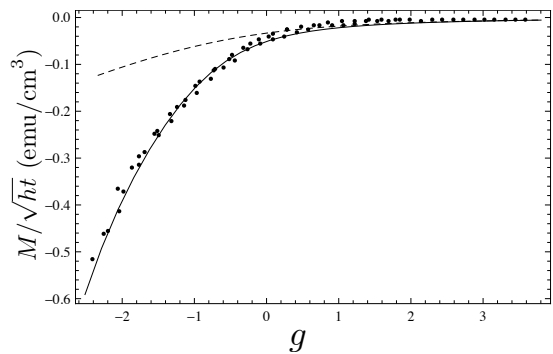}
\caption{Scaled magnetization data 
from Ref.\ \cite{choi09}, at fields 3, 5, and 7 T, 
along with a theoretical fit from Eq.\ \eqref{mag}.  
The theoretical scaling function \eqref{mag} uses fitting parameters 
$D_M = 0.70$ and $\tilde \eta / \alpha_0 = 0.034$ (solid line), 
with other parameters given in the text. The dashed line is the  
2D case ($\tilde \eta = 0$).
\label{mag_fit}}
\end{figure}

Fig.\ \ref{mag_fit} shows fluctuation magnetization data 
\cite{choi09} for $\mathrm{BaFe_{1.8}Co_{0.2}As_2}$,
 and a fit of Eq.\ \eqref{mag} to the data.  
For this sample $T_c(0) = 23.6\ \mathrm K$; and 
we obtain $g_0 = 5.8$ using the values $H_{c2}(0) = 72\ \mathrm T$ for the upper critical field, $\kappa = 44$ for the GL parameter  \cite{williams09}, 
and $s = 6.65\  \mathrm \AA$ for the interlayer 
spacing \cite{rotter08}.  The demagnetization factor $D_M$, 
which reduces the overall magnetization by a 
factor of $1-D_M$, is not known exactly for this sample, 
but can be estimated as $D_M \approx 1 - \pi d / (2 R)$, 
which is valid for a flat disk of radius $R$ and 
thickness $d \ll R$ in a perpendicular magnetic field \cite{fetter67}.  
The sample used in Ref.\ \cite{choi09} is rectangular in shape 
with length and width $L \approx 10d$, so we expect $D_M \approx 1 - \pi / 10$.  
Fitting the data with respect to $D_M$ and 
$\tilde \eta / \alpha_0$, with other parameters fixed, 
yields the curve in Fig.\ \ref{mag_fit} and  
$D_M = 0.70$, $\tilde \eta / \alpha_0 = 0.034$.  

\begin{figure}
\includegraphics[width=0.4\textwidth]{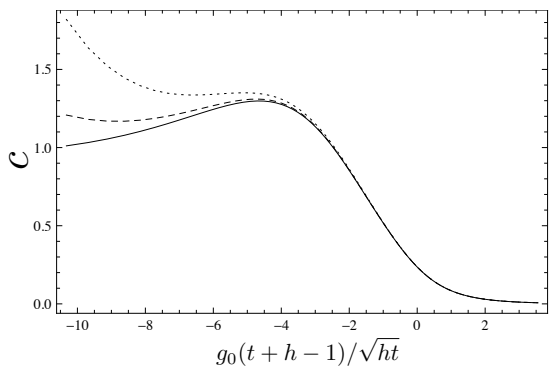}
\caption{Specific heat from Eq.\ \eqref{c_full}, with $g_0 = 3$, $h = 0.3$, and $t_{c2}(h) = 1-h$. The three curves have interlayer coupling values of $\tilde \eta / \alpha_0 = 0$ (solid), 0.002 (dashed), and 0.004 (dotted).\label{sp_heat}}
\end{figure}

We now calculate the heat capacity $C = -T \partial^2 F / \partial T^2$.  From Eq.\ \eqref{eff} we obtain
\begin{eqnarray}
\begin{split}
\label{c_full}
\frac{2 g_0 ^2}{h \beta_\Delta} c  = & \left( 2 \frac{\partial g} {\partial t} + t \frac{\partial ^2 g}{\partial t^2} \right) \left(2 g
U^2 - 2 U \sqrt{2 + g^2 U^2} \right)
\\ & + \frac{1}{2 t} + 2 t \left( \frac{\partial g}{\partial t}
\right)^2 \left(U^2 - \frac{g U^3}{\sqrt{2 + g^2 U^2}} \right) \\ & + 4t \left( \frac{\partial g}{\partial t} \right)^2 \frac{d U}{d g} \left( g U - \frac{1 + g^2 U^2}{\sqrt{2 + g^2 U^2}}\right).
\end{split}
\end{eqnarray}
\noindent 
Here the heat capacity $c \equiv C / \Delta C_{2d}$ has been normalized to its 2D mean-field value, $\Delta C_{2d} = V \alpha_0 ^2 t / (s \beta \beta_\Delta) = 2 V H_{c2}(0) g_0 ^2 t / (\phi_0 s \beta_ \Delta)$, and $g$ is given by \eqref{g}.
Fig.\ \ref{sp_heat} shows $c$
for three different values of $\tilde\eta$.  
As $T\to 0$, there is a divergence in the specific heat,
stemming from the fact that, for $\tilde \eta\not = 0$,
$g \rightarrow - \infty$ at 
{\em finite} $T\to T_{\Delta}$, as discussed before.
This is suggestive of a first-order Abrikosov transition at $T_{\Delta}$;
to describe its details our approach needs to be 
augmented either by the sixth order GL term (since $\beta '\to 0^+$
at $T_{\Delta}$) or finite $d$ corrections,
something left for future study.
The specific heat, being a second derivative, is rather 
sensitive to this divergence at low $T$, even for small $\tilde \eta$, as
we illustrate in the figure.

Recent experiments on $\mathrm{SmFeAsO_{1-x}F_x}$ \cite{pallecchi09}
suggest that the fluctuation conductivity follows an
approximate 2D scaling behavior of the form predicted 
by Ref. \cite{ullah91} (see also Ref. \cite{ikeda}), where transport 
coefficients are derived from the time-dependent GL-LLL theory, 
within the Hartree-Fock approximation ($\beta_A=2$). 
We follow Ref. \cite{ullah91} to obtain the fluctuation 
conductivity  as
\begin{eqnarray}
\label{sigma}
\sqrt{\frac{H}{T}} \Delta \sigma_{yy} = 
\frac{\hbar}{64 \lambda_{ab}(0) \xi_{ab} (0) T_c(0)} 
\sqrt{\frac{\pi}{s \phi_0}} {\cal K} (g),
\end{eqnarray}
\noindent 
where, in their case, the scaling variable $g$ has its
2D form, i.e.\ $\tilde \eta = 0$ in Eq.\ \eqref{g}.  
In obtaining Eq.\ \eqref{sigma}, we used the relations from the GL theory $\alpha_0 =  e^2 H_{c2}^2(0) \lambda_{ab}^2(0) / m_{ab} c^2 \kappa ^2$ and $H_{c2}(0) = \phi_0 / 2 \pi \xi_{ab}^2(0)$, as well as the expression for the coefficient in the time-dependent GL equation $\Gamma_0^{-1} \approx \pi \hbar \alpha_0 / 8 T_c (0)$ \cite{tinkham96,larkin05}.  The scaling function in \eqref{sigma} has 
the form ${\cal K}(g)={\cal K}_{2D} (g) \equiv  -g/2 + \sqrt{1 +  g^2 / 4} $,
where now, of course, the scaling variable $g$ must be changed to our Eq. \eqref{g},
with $\tilde\eta\not =0$.

\begin{figure}
\includegraphics[width=0.4\textwidth]{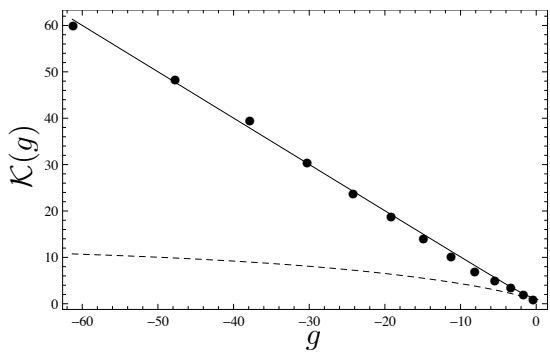}
\caption{Fluctuation conductivity data from Ref.\ \cite{pallecchi09}, 
along with a theoretical fit, with scaling variable
$g$ from \eqref{g}, $g_0=6.37$, and $\tilde \eta / \alpha_0 = 0.022$ (solid line).  
The purely 2D curve ($\tilde \eta = 0$) is shown for comparison (dashed line).  
Other parameters are given in the text.  \label{fluc_cond}}
\end{figure}

Comparison of the scaling function ${\cal K}(g)$ to the
experimental data in \cite{pallecchi09} is not straightforward since 
their sample is a polycrystal. To compensate for this, we replace
$\xi_{ab}(0)\to (\xi_{ab} (0)^2\xi_c(0))^{1/3}$,
$\lambda_{ab}(0)\to (\lambda_{ab} (0)^2\lambda_c(0))^{1/3}$  in 
the prefactor in \eqref{sigma}.
Fig.\ \ref{fluc_cond} shows ${\cal K}(g)$ and the 
data for the optimally doped ($x = 0.15$,  $T_c (0) = 51.5\ \mathrm{K}$) sample at 
$H = 28\ \mathrm T$.
The coherence length is $\xi_{ab(c)}(0) = 24\ (3)\ \mathrm \AA$ \cite{pallecchi09}; 
the penetration depth
$\lambda_{ab(c)}(0) = 2000\ (16000)\ \mathrm{\AA}$ \cite{prozorov09}; 
the upper critical field $H_{c2}(0) / T_c (0) = 7.8\ \mathrm{T/K}$ 
\cite{footnotehighslope} fits snugly between $|dH^{10\%}_{c2}/dT |$ 
and $| dH^{90\%}_{c2}/dT|$ reported in Ref. \cite{pallecchi09};
and the interlayer separation 
$s = 8.45 \mathrm{\AA}$ \cite{margadonna09}.  
One can see that the 
interlayer coupling leads to a strong 
enhancement of conductivity over its 2D form, even 
for modest values of $\tilde \eta / \alpha_0$.

In summary, we showed that a GL theory of
coupled fluctuating superconducting layers in
a magnetic field can be expressed as 
an {\em effective, self-consistent single} layer
problem, in the limit of a large number of neighboring layers.  Our
approach can be generalized to other 2D, 1+1D or 2+1D problems.
Comparison of the theory with experimental results in the iron-pnictides is rather
favorable, and provides a means of making the 
quasi 3D nature of these materials more theoretically tractable.

\begin{acknowledgments}
This work was supported in part by the Gardner Fellowship (JMM)
and the Johns Hopkins-Princeton Institute for Quantum Matter, under Award No.\ DE-FG02-08ER46544 by the U.S.\ Department of Energy, Office of Basic Energy Sciences, Division of Materials
Sciences and Engineering.
\end{acknowledgments}


\bibliographystyle{apsrev}

\begin{thebibliography}{10}
\bibitem{footnotehighslope} Note that such 
a high value of $|dH_{c2}/dT |$ simply reflects
a polycrystalline character of a sample used in Ref. \cite{pallecchi09}, with 
$\xi_c \ll \xi_{ab}(0)$.
\end{thebibliography}

\end {document}